\documentclass[aps,preprint]{revtex4-1}
\usepackage{amsmath,amsfonts,graphicx,epstopdf,hyperref}

\begin{document}

\abovedisplayskip=20pt
\abovedisplayshortskip=20pt
\belowdisplayskip=20pt
\belowdisplayshortskip=20pt

\title{Superoscillations with arbitrary polynomial shape}

\author{Ioannis Chremmos}
\email{ioannis.chremmos@mpl.mpg.de}
\affiliation{Max Planck Institute for the Science of Light, D-91058 Erlangen, Germany}

\author{George Fikioris}
\email{gfiki@ece.ntua.gr}
\affiliation{School of Electrical and Computer Engineering, National Technical University, GR 157-73, Athens, Greece}

\date{\today}

\begin{abstract}
We present a method for constructing superoscillatory functions the superoscillatory part of which approximates a given polynomial with arbitrarily small error in a fixed interval. These functions are obtained as the product of the polynomial with a sufficiently flat, bandlimited envelope function whose Fourier transform has at least $N-1$ continuous derivatives and an $N$-th derivative of bounded variation, $N$ being the order of the polynomial. Polynomials of arbitrarily high order can be approximated if the Fourier transform of the envelope is smooth, i.e. a bump function.
\end{abstract}

\maketitle

\section{Introduction}
Superoscillation is the counter-intuitive property of a bandlimited function to oscillate with local frequencies that are larger than its maximum frequency component. Early explicit reports of this property are found in the field of signals and systems \cite{Sayegh1983, Bucklew1985}, although controlling the oscillations of bandlimited functions through their zeros is a much older subject \cite{Bond1958}. From the viewpoint of physics, superoscillations imply the ability of a radiating or imaging system to produce or resolve wave features much finer than the bandwidth of the system suggests. Elements of the phenomenon can therefore be traced back in the classical quest for superdirective antennas (originating from Einstein's concept of needle radiation) \cite{Oseen1922} and for optical imaging beyond the diffraction-limit \cite{Toraldo1952}. Interestingly, the term \textit{superoscillation} was actually born much later within quantum mechanics as part of the concept of weak measurements \cite{Aharonov1988Spin100, Aharonov1990TimeTranslation}, which subsequently motivated the systematic analysis of \textit{faster-than-Fourier} functions \cite{BerryFasterThanFourier, Berry1994Billiards} and their temporal or spatial evolution as quantum or optical wavefunctions \cite{BerrySuperOscillations}. In recent years, superoscillations enjoyed a renewed interest within the field of super-resolution imaging, as optical technology enabled their use for focusing light in the far field at spatial scales below the diffraction limit \cite{Rogers2012SuperLens, Rogers2013}. In any application involving superoscillations, however, one has to fight against the inevitable diminishing of the signal amplitude in the superoscillatory region, as follows from standard Fourier analysis \cite{Ferreira2006FasterThanNyquist}.

Physical implications aside, the construction of superoscillatory functions is an interesting mathematical problem in itself. Two general approaches can be distinguished in the literature so far. The first uses the sampling theorem to express a superoscillatory function as a series of shifted sinc functions with only a finite number of non-zero coefficients. The latter are determined through a linear system of amplitude constraints on a finite grid of points \cite{Ferreira2007ConstructionAharonov, Lee2014DirectConstruction}. The grid is finer than a Nyquist sampling grid in order to make the function super-oscillate. This approach is also related to the old problem of constructing bandlimited functions with a given set of zeros \cite{Bond1958}. For this particular problem, there is also the possibility to directly replace some of the zeros of a bandlimited function with the desired ones without changing its bandwidth. The zero-replacement theorem has appeared in several independent works \cite{Requicha1980, Bucklew1985}, as well as in explicit connection with superoscillations \cite{Qiao1996}.

The second approach involves expressing the superoscillatory function as a Fourier integral and again imposing a set of amplitude constraints on a fine grid of points. Variational techniques are subsequently used to find the minimum-energy function that solves the problem \cite{Kempf2004UnusualProperties, Ferreira2006FasterThanNyquist}. Constraints for the derivative of the function can be applied as well \cite{Lee2014PrescribedAplitude}. The variational approach too has old roots in the field of information theory \cite{Levi1965}.

In addition to the above methods, the literature provides well-studied examples of superoscillatory functions, such as the familiar $f(t)=(\cos t + i a \sin t)^N$ \cite{Aharonov1988Spin100, BerrySuperOscillations}, as well as specific Fourier-type integral representations where superoscillations emerge due to the presence in the transform of a delta-function-like factor centered in the complex plane \cite{BerryFasterThanFourier}.

A common feature of the existing general methods is that the superoscillatory function is required to satisfy a \textit{discrete} set of constraints, concerning the value of the function or its derivative on a fine grid of points. This however implies limited control over the actual shape of the function in the superoscillatory interval. In sampling methods in particular, the points at which the sinc basis functions are centered can be chosen quite arbitrarily \cite{Ferreira2007ConstructionAharonov} leading to an infinite number of different functions satisfying the same superoscillatory constraints. A similar lack of control over the shape of superoscillations exists when specific methods or prototype functions are used. Their parameters can be tuned to define the maximum local frequency or the total number of the superoscillations but there is little or no flexibility at all in controlling their shape.

An immediate question is whether one can construct superoscillatory functions in a more \textit{continuous} way, namely having control over the shape of the function at least in the superoscillatory region. More specifically, one may ask if it is possible to construct a superoscillatory function $f(t) \in L^2(R)$ (square integrable in the entire real line) that approximates a desired analytic function $p(t)$ with arbitrary accuracy in a certain interval $(-a,a)$. The accuracy of the approximation can be quantified in terms of a norm of the difference $f(t)-p(t)$ over the interval, say $||f-g||_2$, namely
\begin{equation}
\left( \int _{-a}^{a} \left[ f(t) - p(t) \right]^2 dt \right)^{1/2} < \epsilon,
\label{eq:norm}
\end{equation}
for some small $\epsilon$. Of course, for an arbitrary analytic function $p(t)$, one can generally speak only of an approximation because a strict equality $f(t)=p(t)$ over a finite interval would imply, by analyticity, an equality for all $t$ which is generally impossible since $p(t)$ may not even belong in $L^2(R)$, as for example in the case of a polynomial or a trigonometric function.

The answer to the above question is positive. To prove this, we here present a simple method that allows to construct superoscillatory functions that approximate a given polynomial $p_N(t)$ ($N$ being the order) with arbitrarily small error in a finite interval. Although the method refers to polynomials, general analytic functions $p(t)$ can be treated too by first expanding them into their Taylor series around some $t_0 \in (-a,a)$. The expansion is truncated to some order $N$ so that the corresponding Taylor polynomial $p_N(t)$ is an approximation of $p(t)$ in this interval in the sense $||p - p_N||_2 < \epsilon_1$. Then a superoscillatory function $f(t)$ can be found that approximates this polynomial in the sense $||f - p_N||_2 < \epsilon_2$. By the triangle inequality $||f-p||_2 \leq ||f-p_N||_2 + ||p-p_N||_2 < \epsilon_1 + \epsilon_2$ which is a number that can be made smaller than any given $\epsilon$.

\section{Method}
We consider the Paley-Wiener space $PW_{\pi}$ of real square integrable functions $f(t) \in L^2(R)$ whose Fourier transform $F(\omega)$ is supported on $[-\pi,\pi]$, namely bandlimited functions with finite energy and bandwidth $\pi$. We also assume the real polynomial
\begin{equation}
p_N(t) = \sum _{n=0} ^N a_n t^n
\label{eq:polynomial}
\end{equation}
as the target or desired shape of our function in the superoscillatory interval $(-a,a)$. For having at least two zeros in this interval, one requires $N \geq 2$. Now consider a known function $e(t) \in PW_{\pi}$ that we term the \textit{envelope} function. We additionally assume that its Fourier transform $E(\omega)$ is at least $C^{N-1}(-\infty,\infty)$ (i.e. it has at least $N-1$ continuous derivatives for all real $\omega$) and has a $N$-th derivative of bounded variation. Then by the familiar property of Fourier analysis the function $f(t) = p_N(t) e(t)$ has the Fourier transform
\begin{equation}
F(\omega) = \sum _{n=0} ^N a_n i^n E^{(n)}(\omega)
\label{eq:transform}
\end{equation}
where the superscript $(n)$ indicates the $n$-th derivative with respect to the argument. By our assumptions, the above transform is clearly zero for $|\omega| > \pi$ and of bounded variation (hence $L^p$ integrable, $p > 0$), therefore $f(t) \in PW_{\pi}$.

The above can be stated alternatively using the smoothness-decay property of Fourier transforms \cite{TrefethenSpectralMethods}. If the Fourier transform of the envelope function is at least $C^{N-1}(-\infty,\infty)$ and has an $N$-th derivative of bounded variation, then $e(t)$ decays at least as $O(|t|^{N+1})$ as $|t| \to \infty$, so that $p_N(t) e(t)$ is still a function in $PW_{\pi}$, despite the polynomial growth of $p_N(t)$.

Moreover, if $e(0)=1$ and $e(t)$ is sufficiently flat at $t=0$, one has $f(t) \approx p_N(t)$ in the interval of interest $(-a,a)$. The flatness of $e(t)$ can be independently controlled through a dilation transformation $e(t) \to e(t/D)$ where $D \geq 1$. Our superoscillatory function finally reads
\begin{equation}
f(t) = p_N(t) \: e\left( \frac{t}{D} \right) 
\label{eq:function}
\end{equation}
Note that the dilation confines the spectrum of $e(t)$ even more hence $e(t/D)$ is still a member of $PW_{\pi}$.

\section{An example}
\begin{figure}[t]
\includegraphics[width=0.9\textwidth]{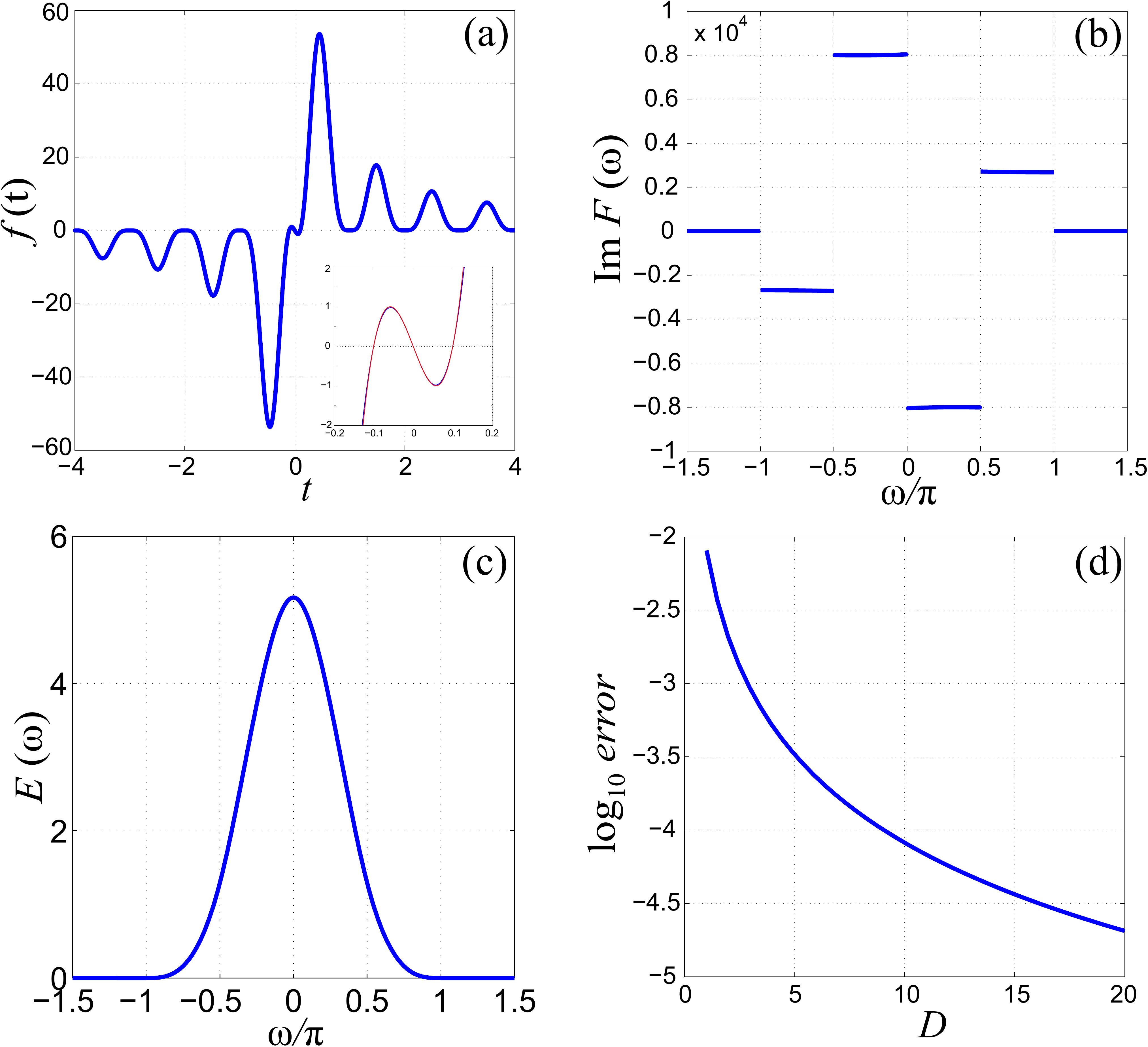}
\caption{(a) The function of Eq. \eqref{eq:example_cubic} for $s=0.1$ and $D=1$. The inset zooms into the superoscillatory interval $(-0.1,0.1)$, also showing the corresponding polynomial $p_3(t)$ in red. (b) The imaginary part of the Fourier transform of $f(t)$ (the real part is zero). (c) The Fourier transform of the envelope function (d) The error $||f-p_3||_2$ in $(-0.1,0.1)$ as a function of the dilation factor $D$.}
\label{fig:example_cubic}
\end{figure}

As an example consider the function
\begin{equation}
f(t) = p_3(t) \: e \left( \frac{t}{D} \right) = \frac{3 \sqrt{3}}{2} \left( \frac{t^3}{s^3} - \frac{t}{s} \right) \: \text{sinc} ^4 \left( \frac{t}{4D} \right) 
\label{eq:example_cubic}
\end{equation}
where $\text{sinc}(x) = \sin(\pi x) / (\pi x)$. It is easy to show, by successive convolutions of the Fourier transform of the sinc function (which is equal to 1 for $|\omega| < \pi$ and zero otherwise), that the function $e(t) = \text{sinc}^4 (t/4)$ has the Fourier transform
\begin{equation}
 E(\omega ) = \int _{-\infty} ^{\infty} e(t) e^{-i \omega t} dt= \frac{16}{\pi ^3} \left\{ {\begin{array}{*{20}c}
   {{\left| \omega  \right|^3}  - \pi \omega ^2 + \frac{\pi^3}{6} ,} & {|\omega | \le \frac{\pi }{2}}  \\ \\
   {\frac{1}{3}\left( \pi  - \left| \omega  \right| \right)^3 ,} & {\frac{\pi }{2} < |\omega | \le \pi }  \\
\end{array}} \right.
\label{eq:example_cubic_envelope_fourier}
\end{equation}
and zero for $|\omega| > \pi$ (plotted in Fig. \ref{fig:example_cubic}(c)). The above is $C^2(-\infty,\infty)$ differentiable with an integrable third derivative hence, according to the previous discussion, $\text{sinc}^4 (t/4)$ is an appropriate envelope to multiply with a cubic polynomial and obtain a $f(t) \in PW_{\pi}$. For $s<1$ and $D \geq 1$, the $f(t)$ of Eq. \eqref{eq:example_cubic} superoscillates in $(-s,s)$ approximately as the cubic polynomial $p_3(t) = \frac{ 3 \sqrt{3} }{2} \left[ (t/s)^3 - t/s \right]$. The prefactor sets the amplitude of the superoscillation to 1.

An example of the function of Eq. \eqref{eq:example_cubic} is shown in Fig. \ref{fig:example_cubic}(a) for $s=0.1$ and $D=1$. Notice that its Fourier transform, shown in Fig. \ref{fig:example_cubic}(b), is hardly distinguishable from the third derivative of the Fourier transform of $\text{sinc}^4(t/4)$. Indeed, according to Eq. \eqref{eq:transform}, we have for our example
\begin{equation}
F(\omega) = -i \frac{3 \sqrt{3}}{2} \left( \frac{1}{s^3}E^{(3)}(\omega) + \frac{1}{s}E^{(1)}(\omega) \right)
\label{eq:example_cubic_fourier}
\end{equation}
which shows that the third derivative term dominates for $s<<1$. However, even in the case of vanishingly small $s$, the first derivative term cannot be neglected as it corresponds to the linear term in the target polynomial $p_3(t)$ which is responsible for its oscillatory (or sinusoidal) shape. This verifies the nature of superoscillations as a delicate interference effect, something that has already been noted in the literature \cite{BerrySuperOscillations} from a different perspective (our present point of view is that of the spectrum).

\section{On the spectrum}
Generalizing the above remark, consider a function $f(t) \in PW_{\pi}$ that superoscillates in $(-a,a)$ with the oscillation amplitude in the order of unity. In the context of our approach, the behaviour of $f(t)$ in $(-a,a)$ can be approximated with the help of a polynomial as for example its truncated Taylor series around $t=0$. If the superoscillations have a characteristic scale $s<<1$, the Taylor polynomial will be of the form
\begin{equation}
p_N(t) = \sum _{n=0} ^N a_n \left( \frac{t}{s} \right) ^n
\label{eq:polynomial_scale_s}
\end{equation}
with the real coefficients $a_n$ being in the order of 1. Stating that alternatively, the magnitude of the derivative $f^{(n)}(0)$ is in the order of $s^{-n}$. This is easily understood by considering polynomials with roots separated by multiples of $s$, for example
\begin{equation}
p_N(t) = \left( \frac{t}{s}-1 \right) \left( \frac{t}{s}-2 \right) ... \left( \frac{t}{s}-N \right)
\label{eq:polynomial_roots_s}
\end{equation}
In the context of our method, the entire superoscillatory function $f(t)$ is the product of such an approximating polynomial with an envelope function $e(t)$ that decays sufficiently fast as $|t| \to \infty$ and is flat enough around $t=0$ with $e(0)=1$. Multiplying Eq. \eqref{eq:polynomial_scale_s} with such an envelope and taking the Fourier transform we obtain
\begin{equation}
F(\omega) = \sum _{n=0} ^N \frac{a_n i^n}{s^n} E^{(n)}(\omega)
\label{eq:transform_scale_s}
\end{equation}
Thus $F(\omega)$ contains contributions of the derivatives of $E(\omega)$ weighted with $s^{-n}$ which makes the contribution of the highest-order derivative $E^{(N)}(\omega)$ dominant. Nevertheless, despite the spectrum of $f(t)$ resembling $E^{(N)}(\omega)$, its superoscillatory behaviour is actually due to the progressively weaker, essentially perturbative, contributions of the lower-order derivatives $E^{(n)}(\omega)$ with $n<N$. This explains why it is difficult to tell if a bandlimited function is superoscillatory from its spectrum (``there is no hint of superoscillations in the power spectrum'' \cite{BerrySuperOscillations}).

The smallness of the amplitude of superoscillations can also be quantified in the context of our approach. Having normalized their amplitude to unity, the amplitude of the function increases outside the interval $(-a,a)$ according to the approximating polynomial $p_N(t)$. The maximum is reached at some $t \sim D$ and, according to Eq. \eqref{eq:polynomial_roots_s}, is in the order of $(D/s)^N$. Since $N$ determines the number of zeros in the superoscillatory interval, we have verified the exponential dependence of the amplitude on the number of superoscillations \cite{Ferreira2006FasterThanNyquist}.

\section{Conclusion}
We have presented a simple method for designing superoscillatory functions that approximate a given polynomial $p_N(t)$ with arbitrarily low error within a given interval. The functions are obtained by multiplying the polynomial with a sufficiently flat bandlimited envelope function $e(t)$ whose Fourier transform is (at least) as smooth as needed to keep the product in the Payley-Wiener space. Appropriate envelope functions can be obtained, for example, as powers of the sinc function or, more generally, as the Fourier transform of polynomial splines. Envelope functions that work with polynomials of arbitrarily high order $N$ can also be constructed. Such are the inverse Fourier transforms of bump functions, namely $C^{\infty}(-\infty,\infty)$ functions with a compact support, as for example $E(\omega) = e^{-\frac{1}{\pi^2-\omega^2}}$ for $|\omega| < \pi$ and $E(\omega)=0$ for $|\omega| \geq \pi$. We have finally seen that the superoscillatory nature of a bandlimited function constructed with present method is hidden in its spectrum as a series of geometrically diminishing contributions from derivatives of the envelope's Fourier transform.


\end{document}